# Electric-field Switching of Interlayer Magnetic Order in a van der Waals Heterobilayer via Spin-potential Coupling

Chengxi Huang[1], Jinzhe Han[1], Jing Wang[1], Jintao Jiang[1], Ziyang Qu[1], Fang Wu[2], Ang Li[1], Yi Wan[1], Kaiyou Wang[3,4*], Erjun Kan[1*]

[1] *MIIT Key Laboratory of Semiconductor Microstructure and Quantum Sensing, Nanjing University of Science and Technology, Nanjing 210094, P. R. China*

[2] *College of Information Science and Technology, Nanjing Forestry University, Nanjing, Jiangsu 210037, P. R. China.*

[3] *State Key Laboratory for Superlattices and Microstructures, Institute of Semiconductors, Chinese Academy of Sciences, Beijing 100083, P. R. China*

[4] *Center of Materials Science and Optoelectronics Engineering, University of Chinese Academy of Sciences, Beijing 100049, China*

\* Correspondence and requests for materials should be addressed to
E. K. (ekan@njust.edu.cn), K. W. (kywang@semi.ac.cn)

**Abstract**

Electric-field switching of magnetic order is of significant physical interest and holds great potential for spintronic applications. However, it has rarely been reported in two-dimensional (2D) van der Waals (vdW) magnets due to the inherently weak interaction between spin order and electric fields. Here we propose a general spin-potential mechanism that significantly enhances the magnetoelectric coupling. As a result, the relative stability of different interlayer magnetic orders in an asymmetric van der Waals heterobilayer can be reversed by external electric fields via spin-potential coupling. Based on this mechanism, we designed a series of 2D vdW all-magnetic heterobilayers, such as $CrI_3$/$MnSe_2$, in which a transition from interlayer ferromagnetic (iFM) to antiferromagnetic (iAFM) order is realized by a feasible electric field around 0.1 V/Å. Our findings not only reveal a novel magnetoelectric coupling mechanism, but also present a practical strategy for achieving pure electric-field-driven magnetic order switching.

**Introduction**

Modern magnetic memories primarily rely on a fundamental physical effect that the magnetoresistance of a magnetic tunnel junction (MTJ) can be significantly changed by switching its magnetic state. Controlling the magnetic state by an external electric field (voltage), rather than a current or magnetic field, has been a long-sought goal for developing next-generation, low-dissipation and high-efficiency spintronic devices, such as voltage-controlled magnetic random access memories[1,2]. Interestingly, a giant tunnelling magnetoresistance has also been observed in two-dimensional (2D) van der Waals (vdW) magnetic layers[3,4,5], since the discovery of 2D magnets[6,7,8,9,10,11,12]. Such a magnetoresistance can be significantly modulated by altering the interlayer magnetic state through various external factors, such as magnetic fields[13], mixed magnetic and electric fields[14,15,16], strain[17], pressure[18,19], twisting[20] and spin injection[21]. These advancements have opened up a new avenue toward vdW spin-filter MJT[22]. However, pure electric-field switching of interlayer magnetic order remains a significant challenge owing to the lack of effective magnetoelectric coupling mechanism.

Generally, magnetoelectric coupling could be induced through spin-orbit, spin-lattice and/or spin-charge interactions[23]. Although plenty of 2D vdW magnets have been discovered so far, electrical switching between interlayer ferromagnetic (iFM) and antiferromagnetic (iAFM) orders has only been reported in the $CrI_3$ bilayer[24,25,26], driven by the charge transfer mechanism. As shown in Fig. 1a, extra electron or hole carriers transfer from the substrate to the vdW magnetic layers upon electrostatic gating, which increases the density of states at the Fermi level and typically leads to an iAFM-to-iFM order transition[27,28]. However, the spin-charge interaction induced by interlayer charge transfer is commonly rather weak, so that the switching of interlayer magnetic order still requires the assistance of a strong external magnetic field. Therefore, a general mechanism to enhance the coupling between spin order and external electric field is urgently needed for achieving pure electric-field-driven interlayer magnetic order switching in 2D vdW magnets.

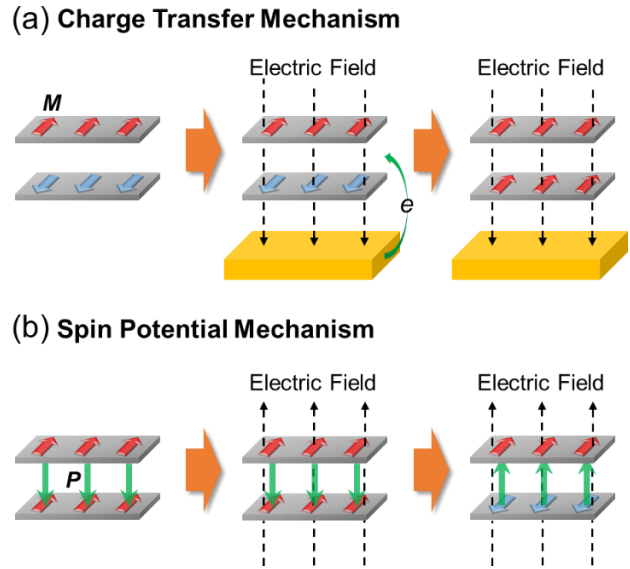

**Figure 1 | Interfacial magnetoelectric coupling mechanism.** Schematic diagrams for electric-field switching of interlayer magnetic order driven by (a) charge transfer and (b) spin potential mechanism. Gray and yellow slabs represent magnetic vdW monolayers and non-magnetic substrates, respectively. Red and Blue arrows represent spin moments. Green arrows represent electric dipole moments.

Here we propose a general spin-potential coupling mechanism that in an asymmetric vdW magnetic bilayer, the interlayer polar spin interaction could lead to distinct different out-of-plane electric dipole moments for different interlayer spin orders. Consequently, under an external out-of-plane electric field, the iFM and iAFM orders gain very different electric dipole potential energies, which could reverse their relative stability (Fig. 1b). Based on a tight-binding dimer model, we reveal that the remarkable interplay between spin order and electric dipole moment originates from the disparity in occupation and spatial distribution of spin charges between ferromagnetic (FM) and antiferromagnetic (AFM) states. Then utilizing first-principles calculations, we predict a series of asymmetric vdW magnetic heterobilayers that exibits strong magnetoelectric coupling driven by the spin-potential mechanism. Especially, for the $CrI_3$/$MnSe_2$ heterobilayer, an iFM-to-iAFM phase transition could be realized by applying a feasible out-of-plane electric field around 0.1 V/Å.

**Results**

**The concept of spin potential.** It is known that the electric polarization of a ferroelectric material can be manipulated by external electric fields because of the dipole potential energy determined by the electric dipole moment (Fig. 2a). Similarly, the magnetic order of a magnetic material can be controlled by external magnetic fields owing to the magnetic potential energy relying on magnetic moment. Theoretically, there is no direct interaction between a spin moment and an electrostatic field. Consequently, magnetic order is hardly affected by external electric fields, unless there exist an electric potential energy depending on spin orders. Let's assume that the FM and AFM states of an asymmetric magnetic dimer have different electric dipole moment, so that we can define a parameter $P_{ex} = P_{AFM} - P_{FM} \neq 0$ ($P_{AFM}$ and $P_{FM}$ are electric dipole moments for AFM and FM states). Then according to the definition of electric dipole potential energy $W = -P \cdot \mathcal{E}$ (here $P$ and $\mathcal{E}$ refer to electric dipole moment and electric field intensity, respectively), the electric potential energy difference between FM and AFM orders can be written as $\Delta W = W_{AFM} - W_{FM} = -(P_{AFM} - P_{FM}) \cdot \mathcal{E} = -P_{ex} \cdot \mathcal{E}$. Such a spin-order-dependent additional potential energy is denoted as spin potential, which can be written as $W_{spin} = -P_{spin} \cdot \mathcal{E}$, where $P_{spin}$ represent the electric dipole moment induced by spin interaction. The strength of spin potential for any possible spin order is reflected by the magnitude of $P_{ex}$.

The relative stability of FM and AFM states is determined by their energy difference, defined as $E_{ex} = E_{AFM} - E_{FM}$. Apparently, the different spin potentials for FM and AFM orders will cause a change of $E_{ex}$, namely $\Delta E_{ex}$, by varying the external electric field. Without loss of generality, here we omit the electric field effect on the $P_{ex}$ and obtain $\Delta E_{ex}/\Delta \mathcal{E} = -P_{ex}$, where $\Delta E_{ex}/\Delta \mathcal{E}$ represents the response of $E_{ex}$ to the external electric field. In a vdW magnetic bilayer, the interlayer magnetic order can be considered as a summation of spin dimers consisting of two nearest-neighbouring magnetic ions from different layers. Therefore, the equation $\Delta E_{ex}/\Delta \mathcal{E}_z = -P_{ex}$ (here $E_{ex} = E_{iAFM} - E_{iFM}$, $P_{ex} = P_{iAFM} - P_{iFM}$ and $\mathcal{E}_z$ represents out-of-plane electric field) should also hold for a vdW magnetic bilayer. This suggests that, as long as the magnetic bilayer has a sizable $P_{ex}$, the $E_{ex}$ and, thus, the relative stability of iFM and iAFM orders could be significantly manipulated by applying an out-of-plane electric field (Fig. 2b). However, the $P_{ex}$ is usually ignored in a magnetic system. Because the ionic displacement and/or charge redistribution, which are common origins of electric dipole moments, caused by changing the spin order are generally quite small. Whether and how an electric dipole moment could be greatly affected by switching the spin order is still unclear.

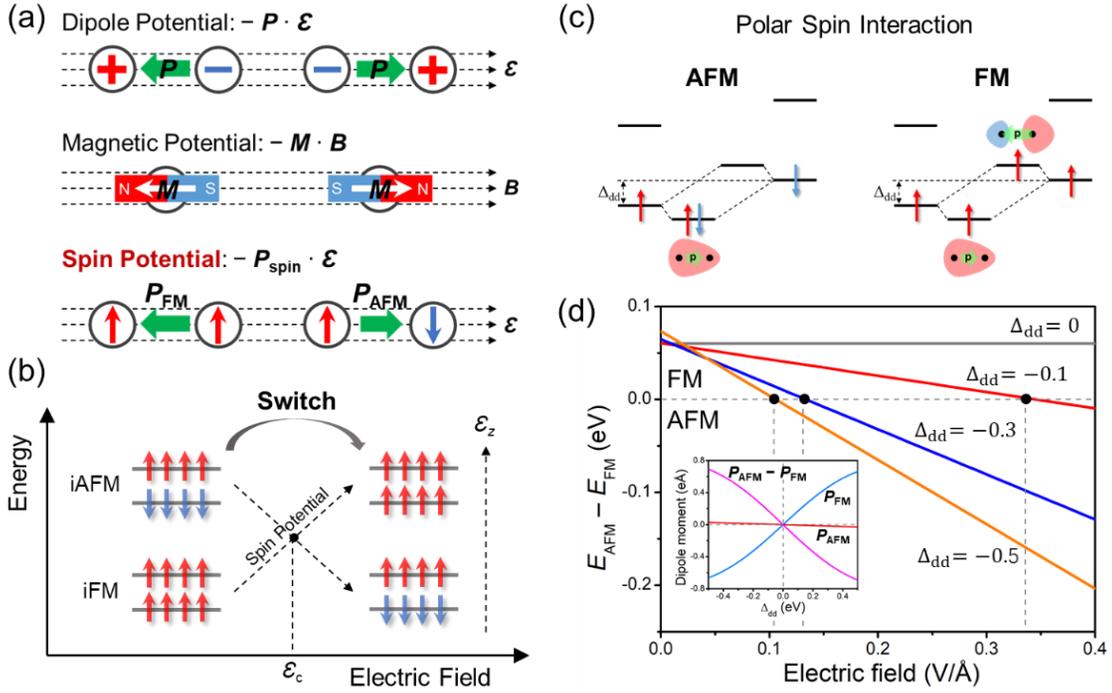

**Figure 2 | Spin potential and driven magnetoelectric coupling mechanism.** (a) Schematic diagram of spin potential. *P*, *M*, *Ɛ* and *B* represent electric dipole moment, magnetic moment, electric field and magnetic field, respectively. $P_{spin}$ represent electric dipole moment induced by spin interaction. Red and blue arrows represent electron spins. Green arrows represent electric dipole moments. (b) Schematic diagram of electric-field switching of interlayer magnetic order driven by spin potential. (c) Tight-binding dimer model for the origin of spin-order-dependent electric dipole moment. $\Delta_{dd}$ represents the on-site energy difference between magnetic atoms. (d) Energy difference between FM and AFM orders ($E_{ex} = E_{AFM} - E_{FM}$) as a function of external electric fields with $\Delta_{dd}$ = 0, −0.1, −0.3 and −0.5 eV derived from the tight-binding model. Inset: electric dipole moments of AFM ($P_{AFM}$), FM ($P_{FM}$) states and their difference ($P_{ex} = P_{AFM} - P_{FM}$) for the magnetic dimer as a function of $\Delta_{dd}$.

**Tight-binding dimer model.** To answer these questions, we firstly introduce a tight-binding dimer model[29] to explore the interplay between electric dipole moment and spin order. For simplicity, we adopted the one-dimensional Gaussian functions $[Ne^{-(x+x_0)^2}$, where $x_0$ represents the position of each atom] as the atomic basis functions to calculate the electron density and electric dipole moment. The general Hamiltonian of the dimer is written as

$$\hat{H} = \sum_{li} \epsilon_{li} \hat{d}^\dagger_{li} \hat{d}_{li} + \sum_{\alpha i, \beta j} [t_{\alpha i,\beta j} \hat{d}^\dagger_{\alpha i} \hat{d}_{\beta j} + h.c.] + \frac{U}{2} \sum_l \vec{e}_l \cdot \vec{S}_l,$$

where $l = \alpha, \beta$ is the site index, $i$ and $j$ are orbital index. The first, second and third terms represent the on-site energy of each orbital, the interatomic hopping and the spin-splitting under the mean field approximation, respectively (see Supplementary section 1 for details).

In the magnetic dimer, each atom has two orbitals and one localized spin (Fig. 2c). If we introduce an on-site orbital energy difference ($\Delta_{dd}$) between the two magnetic atoms, the polar spin interaction will induce an electric dipole moment, which is totally contributed by the spatial disproportion of spin charges and, thus, could be sensitive to the spin order. In the AFM order, the occupied eigenstates are both bonding states (Supplementary Fig. 2d). Whereas in the FM order, the lowest eigenstate is a bonding state while the second-lowest eigenstate is an antibonding state (Supplementary Fig. 2e). This disparity in occupation states leads to a significant redistribution of spin charges and, thus, induces a distinct $P_{ex}$. The magnitude of $P_{ex}$ is positively correlated with the magnitude of $\Delta_{dd}$ (inset in Fig. 2d). Consequently, the $E_{ex}$ is more sensitive to the electric field when the $\Delta_{dd}$ is larger (Fig. 2d) according to the equation $\Delta E_{ex}/\Delta \mathcal{E}_z = -P_{ex}$. Furthermore, if we ignore the electric field effects on the $\Delta_{dd}$ and $P_{ex}$, then we obtain $\mathcal{E}_c = E^0_{ex}/P_{ex}$, where $\mathcal{E}_c$ is defined as the critical $\mathcal{E}_z$ that makes $E_{ex}$ become zero and $E_{ex}^0$ is the $E_{ex}$ under zero field. Supplementary Fig. 5b shows that the $E_{ex}^0$ is only slightly affected by $\Delta_{dd}$. Therefore, the magnitude of $\mathcal{E}_c$ could be reduced by increasing the magnitude of $\Delta_{dd}$ (Fig. 2d and Supplementary Fig. 5c). On the other hand, the $P_{ex}$, $E_{ex}^0$, and $\mathcal{E}_c$ are also affected by the interatomic hopping integral ($t$) (Supplementary Fig. 2c and Fig. 5). Overall, the model analysis results indicate that the $P_{ex}$ and spin potential originate from the polar spin interaction in an asymmetric magnetic system with nonzero $\Delta_{dd}$.

**Magnetic heterodimer.** One alternative way to obtain an intrinsic large $\Delta_{dd}$ in a realistic magnetic system is choosing different transition metal atoms to construct a magnetic heterostructure. The simplest magnetic heterostructure is the transition metal heterodimer, e.g. MnCr dimer. Our first-principles calculation results (Supplementary Fig. 6) show that, for the MnCr dimer, the $P_{AFM}$ is indeed distinctly different from the $P_{FM}$, resulting in a large $P_{ex}$ (> 0.06 eÅ) when the interatomic distance ($r$) is ranging from 3.2 to 4.2 Å. Generally, the interatomic hopping integral $t \propto e^{-r}$ ($e$ is the Euler's number). Therefore, the calculation results (Supplementary Fig. 6) are consistent with our model analysis results (Supplementary Fig. 2c).

**vdW magnetic heterobilayer.** A vdW all-magnetic heterostructure, in which each layer has different magnetic transition metal atom, may also possess a sizable $\Delta_{dd}$. Therefore, here we focus on a vdW all-magnetic heterobilayer, i.e. $CrI_3/MnSe_2$, to explore the interplay between out-of-plane electric dipole moment and interlayer spin order. The $CrI_3$ (Fig. 3a) and $MnSe_2$ (Fig. 3b) monolayers have both been predicted and experimentally confirmed to be 2D FM materials[7,10]. Individually, they are both centrosymmetric and non-polar. Since the lateral lattice constants of $CrI_3$ (6.98 Å) are much larger than that of $MnSe_2$ (3.61 Å), we used a 2×2 supercell of $MnSe_2$ to construct the $CrI_3/MnSe_2$ heterobilayer (with a lattice mismatch of ~3.4%). Fig. 3c shows the optimal stacking structure for both iFM and iAFM orders (see

Supplementary section 2 for details), which has in-plane $c_3$ symmetry and, thus, an out-of-plane electric polarization. The $P_{iAFM}$ and $P_{iFM}$ are calculated to be 0.0202 and −0.0073 eÅ/u.c., respectively, with positive value indicates the direction pointing from MnSe$_2$ to CrI$_3$. The $P_{ex}$ is 0.0275 eÅ/u.c.. Note that, although the system is metallic in the periodic (in-plane) direction, electric polarization along the non-periodic (out-of-plane) direction is well-defined and is tunable by applying an $E_z$, as demonstrated in metallic WTe$_2$ multilayers[30].

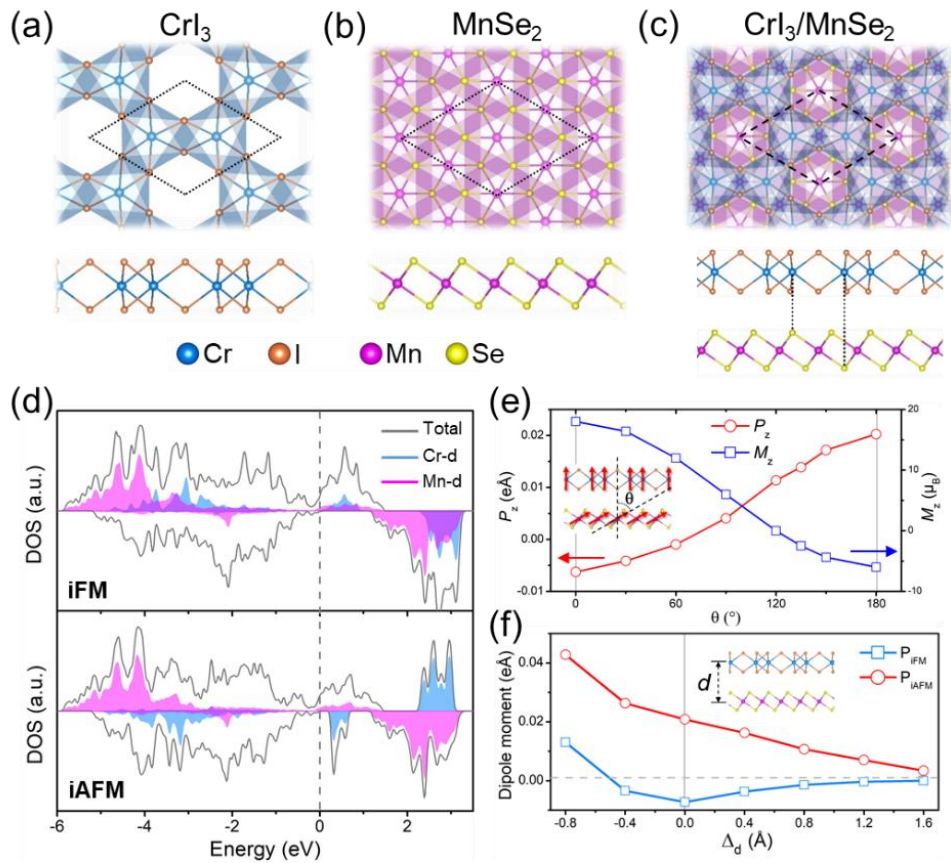

**Figure 3 | CrI$_3$/MnSe$_2$ vdW heterobilayer.** Top (upper panels) and side (lower panels) views of the optimized atomic structures for (a) CrI$_3$, (b) MnSe$_2$ monolayers and (c) CrI$_3$/MnSe$_2$ heterobilayer. Blue, orange, purple and yellow balls represent Cr, I, Mn and Se ions, respectively. (d) Projected density of states for the iFM and iAFM

states of the CrI$_3$/MnSe$_2$ heterobilayer. Gray dashed line represents the Fermi level. (e) Out-of-plane electric dipole moment ($P_z$) and magnetic moment ($M_z$) per unit cell as a function of the angle (θ) between the spins in the two magnetic layers for. (f) The $P_{iFM}$ and $P_{iAFM}$ as a function of the change ($\Delta_d$) of interlayer distance with respect to the equilibrium state.

To understand the origin of such a large $P_{ex}$ for CrI$_3$/MnSe$_2$, we firstly compare the atomic structures of iAFM and iFM states and find that the structural difference is quite small (see Supplementary section 2 for details). Even though we use the same atomic structure for iAFM and iFM states without optimization, the obtained $P_{iAFM}$, $P_{iFM}$ and $P_{ex}$ are 0.0216, −0.0056 and 0.0272 eÅ/u.c., respectively, very close to those of the optimized CrI$_3$/MnSe$_2$. Therefore, the small structural distortion (ionic polarization) is not the main origin of the large $P_{ex}$. Next we investigate the electronic polarization owing to the interlayer charge redistribution, which can be clearly observed in the charge difference between the bilayer and the individual monolayers (Supplementary Fig. 8). The charge accumulation in the interlayer space implies a strong interlayer covalent interactions dominated by Se and I ions on the inner surfaces. From the Hamiltonian matrix elements in the maximally localized Wannier functions (MLWFs) basis (Supplementary Fig. 11), the interlayer interaction between Se-p$_z$ and I-p$_z$ orbitals is comparable to the intralayer interactions. These could lead to a sizable indirect interaction between magnetic Cr and Mn ions. On the other hand, Fig. 3d shows a distinct energy difference between Cr-3d and Mn-3d orbitals, implying a large $\Delta_{dd}$. From the MLWFs results, we obtain a $\Delta_{dd}$ between Cr-3d and Mn-3d levels of ~1.41 eV. Therefore, the large $P_{ex}$ should originate from the interlayer spin charge redistribution due to the sizable $\Delta_{dd}$ and $t$ between magnetic Cr and Mn ions, as revealed in the dimer model.

The origin of $P_{ex}$ and its relation with $\Delta_{dd}$ and $t$ could also be reflected in Supplementary table 2, which includes calculation results of a series of vdW all-magnetic bilayers. First of all, the emergence of nonzero $P_{ex}$ requires the breaking of spatial inversion symmetry via asymmetric interlayer stacking or constructing a heterobilayer. The $P_{ex}$ for individual CrI$_3$ and MnSe$_2$ bilayers are both zero because of their structural centrosymmetry. Secondly, the magnitude of $P_{ex}$ is affected by the type of magnetic transition metal (TM) atoms in different layers. For instance, among the CrI$_3$/TMSe$_2$ (TM = Cr, Mn, Fe) bilayers, the CrI$_3$/CrSe$_2$ has the smallest $P_{ex}$ (0.0015 eÅ/u.c.) because the $\Delta_{dd}$ between the same TM atoms (Cr) in different layers should be small. Thirdly, the choice of ligand ions also affects the $P_{ex}$. For instance, the CrI$_3$/MnTe$_2$ (0.0713 eÅ/u.c.) has much larger $P_{ex}$ than the CrI$_3$/MnSe$_2$ (0.0275 eÅ/u.c.) does. This is because Te has stronger covalency than Se, so that the interfacial Te-I interactions are stronger, leading to a larger indirect $t$ between Cr and Mn atoms and, thus, a larger $P_{ex}$. Similar behaviour also occurs in the CrGeTe$_3$/MnX$_2$ (X = S, Se, Te) systems.

To directly show the magnetoelectric effect in CrI$_3$/MnSe$_2$, we rotate the spins in MnSe$_2$ layer by an angle (θ) ranging from 0 to 180° while the spin direction in CrI$_3$ layer is fixed (θ = 0 and 180° corresponding to iFM and iAFM states, respectively). Fig. 3e shows a continuous change of out-of-plane electric dipole moment (from −0.0073 to 0.0202 eÅ/u.c.) and total magnetic moment along z-axis (from 18 to −6 μ$_B$/u.c.) with respect to the θ, indicating a strong interfacial magnetoelectric coupling.

The relation between interlayer distance and magnetoelectric property of the CrI$_3$/MnSe$_2$ has also been investigated. Supplementary Fig. 12 shows a Bethe–Slater-curve-like relation between $E_{ex}^0$ and interlayer distance, which has also been observed in other magnetic bilayers (e.g. CrSe$_2$)[31]. Fig. 3f shows that the $P_{iAFM}$ monotonously increases, while the $P_{iFM}$ first decreases and then increases as the interlayer distance decreasing. These profiles can also be reproduced by our dimer model by combining the contribution of non-magnetic and magnetic atoms to the electric dipole moments (Supplementary Fig. 2e).

**Electric field effect on magnetic order.** Now we explore the response of $E_{ex}$ to $\mathcal{E}_z$ and its relation with $P_{ex}$ for different vdW magnetic bilayers. As shown in Fig. 4a and Supplementary Fig. 13, the $\Delta E_{ex}/\Delta \mathcal{E}_z$, namely the slope of the profiles, is distinctly different for different magnetic bilayers and is closely related with the $P_{ex}$. For instance, when the $\mathcal{E}_z$ increases from 0 to 0.2 V/Å (positive value indicates the direction pointing from MnSe$_2$ to CrI$_3$), the $\Delta E_{ex}$ of CrI$_3$ bilayer is only 0.14 meV, while that of CrI$_3$/MnSe$_2$ is as large as −6.91 meV. Correspondingly, the $P_{ex}$ of CrI$_3$ bilayer and CrI$_3$/MnSe$_2$ are zero and 0.0275 eÅ/u.c., respectively. Therefore, a sizable $P_{ex}$ can significantly increase the magnitude of $\Delta E_{ex}/\Delta \mathcal{E}_z$. Besides, for a system with zero $P_{ex}$, the $E_{ex}$ for a negative and a positive $\mathcal{E}_z$ are the same (see the inset in Fig. 4a). While those are distinctly different when the $P_{ex}$ is non-zero. More interestingly, by fitting the calculation results of a series of vdW magnetic heterobilayers, we obtained a nearly linear relationship between $\Delta E_{ex}/\Delta \mathcal{E}_z$ and $P_{ex}$ (Fig. 4b), namely $\Delta E_{ex}/\Delta \mathcal{E}_z \approx -P_{ex}$, which demonstrates our aforementioned hypothesis. These indicate that our proposed mechanism is generally applicable and has a dominant effect on the magnetoelectric response for a vdW magnetic system with a sizable $P_{ex}$.

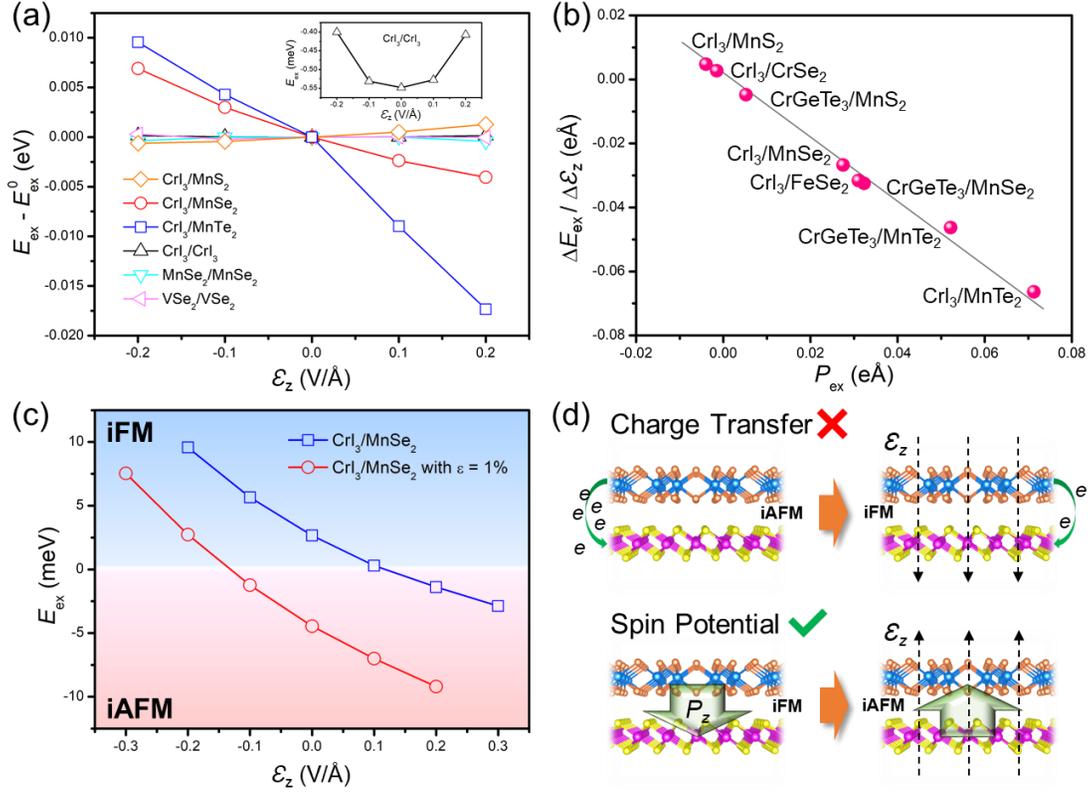

**Figure 4 | Electric field effects on interlayer magnetic coupling.** (a) The change of $E_{ex}$ as a function of $\mathcal{E}_z$ for different vdW magnetic bilayers. A positive value of $\mathcal{E}_z$ indicates the out-of-plane direction pointing from $MnX_2$ (X = S, Se, Te) to $CrI_3$ layer. (b) The nearly linear relation between $\Delta E_{ex}/\Delta \mathcal{E}_z$ and $P_{ex}$. Here, $\Delta E_{ex}/\Delta \mathcal{E}_z = (E_{ex}^{0.1} - E_{ex}^{-0.1})/0.2$, where $E_{ex}^{0.1}$ and $E_{ex}^{-0.1}$ represent the $E_{ex}$ with $\mathcal{E}_z$ = 0.1 and −0.1 V/Å. The gray line indicates the linear fitting of the results. (c) $E_{ex}$ as a function of $\mathcal{E}_z$ for $CrI_3/MnSe_2$ bilayer without and with an in-plane biaxial tensile strain of 1%. (d) Comparison between charge transfer and spin potential mechanism in the $CrI_3/MnSe_2$

The $E_{ex}^0$ is also an important parameter for realizing electric-field-driven magnetic order switching. Supplementary table 2 shows that the magnitude of $E_{ex}^0$ for a vdW magnet is usually ranging from several to dozens of meV/u.c.. Such a large $E_{ex}^0$ makes the electric-field, even magnetic-field control of interlayer spin order very difficult. However, according to our proposed equation $\mathcal{E}_c = E_{ex}^0/P_{ex}$, as long as the $P_{ex}$ reaches several 0.01 eÅ/u.c., the $\mathcal{E}_c$ could be reduced to the order of magnitude of 0.1 V/Å, which is feasible in experiments. Besides, we find that the $E_{ex}^0$ is usually sensitive to interlayer stacking[19,32,33] (Supplementary Fig. 7b) and in-plane strain[17,34] (Supplementary Fig. 14), while the $P_{ex}$ is not. For instance, for the CrI$_3$/MnSe$_2$, the $E_{ex}^0$ greatly changes from 13.10 to −35.85 meV/u.c., while the $P_{ex}$ slightly changes from 0.0201 to 0.0298 eÅ/u.c. when the in-plane strain increases from −4% to 4%. Therefore, the $\mathcal{E}_c$ can also be manipulated by controlling the interlayer stacking or applying an external in-plane strain.

Owing to the large $P_{ex}$ of 0.0275 eÅ/u.c., the $E_{ex}$ of the CrI$_3$/MnSe$_2$ is very sensitive to the $\mathcal{E}_z$ (Fig. 4c). A $\mathcal{E}_z$ greater than ∼0.1 V/Å will change its $E_{ex}$ from 2.66 meV/u.c. to a negative value and, thus, switch the interlayer spin order from iFM to iAFM. The relation between these values is consistent with the equation $\mathcal{E}_c = E_{ex}^0/P_{ex}$ = 0.00266eV/0.0275eÅ ≈ 0.1 V/Å. This mechanism can cause not only an iFM-to-iAFM, but also an iAFM-to-iFM phase transition. For instance, a tensile in-plane strain of 1% changes the interlayer spin order of CrI$_3$/MnSe$_2$ to iAFM with $E_{ex}^0$ = −4.46 meV/u.c while the $P_{ex}$ barely changes. Then a negative $\mathcal{E}_z$ greater than 0.16 V/Å in magnitude could switch the interlayer spin order from iAFM to iFM (Fig. 4c).

It is worth noting that charge transfer mechanism cannot explain the electric-field switching of interlayer magnetic order in the CrI$_3$/MnSe$_2$. Because the Bader charge analysis results show that the intrinsic interlayer charge transfer are −0.0184 and −0.0029 e/u.c. for iAFM and iFM orders, respectively, with negative values indicate a net electron transfer from CrI$_3$ to MnSe$_2$ layer (Fig. 4d). In this case, applying a negative $\mathcal{E}_z$ suppresses the interlayer charge transfer, but enhances the iFM coupling. This violates the aforementioned charge transfer mechanism. Therefore, our proposed spin potential mechanism provides a new angle of view to understand the electric-field switching of magnetic order.

**Conclusions**

In summary, utilizing a tight-binding dimer model and first-principles calculations, we have proposed a general mechanism of realizing electric-field switching of interlayer magnetic order in a vdW magnetic system by introducing a new concept of spin potential and an important parameter $P_{ex}$. The origin of $P_{ex}$ is revealed to be the polar spin interaction in an asymmetric magnetic structure with non-zero $\Delta_{dd}$. The polar spin interaction causes a distinct difference on occupation and spatial distribution of spin charges between iAFM and iFM states. The response of $E_{ex}$ to $\mathcal{E}_z$, which represents the strength of magnetoelectric coupling, is basically proportional to the magnitude of $P_{ex}$, namely $\Delta E_{ex}/\Delta \mathcal{E}_z = -P_{ex}$, and the $\mathcal{E}_c$ is also closely related with the $P_{ex}$, namely $\mathcal{E}_c = E_{ex}^0/P_{ex}$. Consequently, by inducing a sizable $P_{ex}$ via constructing a vdW all-magnetic heterostructure, the relative stability of iAFM and iFM orders can be significantly manipulated by applying an $\mathcal{E}_z$, leading to an electric-field-driven interlayer magnetic order switching, as demonstrated in the CrI$_3$/MnSe$_2$ and other vdW magnetic heterobilayers. These findings will be of great interest for developing pure voltage-controlled vdW spin-filter MTJs. We look forward to experimental observation in the future.

**Methods**

**Computational methods.** The first-principles calculations based on density functional theory were carried out using the Vienna Ab initio Simulation Package[35]. The PBE+U method[36,37] with U$_{eff}$ = 3 eV for 3d electrons of transition metal atoms was used to treat the exchange-correlation and strong electronic correlation. The energy cutoff for the plane wave[38] was 500 eV. A Γ-centered 12 × 12 × 1 Monkhorst-Pack[39] grid was used to sample the Brillouin zone. The vdW correction[40] was included. To avoid the incorrect periodic interactions along the c axis, a vacuum slab of ~30 Å and the dipole-dipole interaction corrections[41] were adopted. The energy and force convergence criteria were 1×10$^{-6}$ eV and 0.005 eV/Å, respectively. The out-of-plane electric dipole moment was calculated by integrating the electron density times the coordination over the whole unit cell.

**References:**

1. Fert, A., Ramesh, R., Garcia, V., Casanova, F., & Bibes, M. Electrical control of magnetism by electric field and current-induced torques. *Rev. Mod. Phys.* **96**, 015005 (2024).
2. Matsukura, F., Tokura, Y., & Ohno, H. Control of magnetism by electric fields. *Nat. Nanotechnol.* **10**, 209–220 (2015).
3. Song, T. et al. Giant tunneling magnetoresistance in spin-filter van der Waals heterostructures. *Science* **360**, 1214-1218 (2018).
4. Zhu, W. et al. Large Tunneling Magnetoresistance in van der Waals Ferromagnet/Semiconductor Heterojunctions. *Adv. Mater.* **33**, 2104658 (2021)
5. Zhu, W. et al. Large Room-Temperature Magnetoresistance in van der Waals Ferromagnet/Semiconductor Junctions. *Chin. Phys. Lett.* **39**, 128501 (2022).
6. Gong, C. et al. Discovery of intrinsic ferromagnetism in two-dimensional van der Waals crystals. *Nature* **546**, 265–269 (2017).
7. Huang, B. et al. Layer-dependent ferromagnetism in a van der Waals crystal down to the monolayer limit. *Nature* **546**, 270–273 (2017).
8. Deng, Y. et al. Gate-tunable room-temperature ferromagnetism in two-dimensional Fe$_3$GeTe$_2$. *Nature* **563**, 94–99 (2018).
9. Bonilla, M. et al. Strong room-temperature ferromagnetism in VSe$_2$ monolayers on van der Waals substrates. *Nat. Nanotechnol.* **13**, 289–293 (2018).
10. O'Hara, D. J. et al. Room Temperature Intrinsic Ferromagnetism in Epitaxial

Manganese Selenide Films in the Monolayer Limit. *Nano Lett.* **18**, 3125–3131 (2018).

11. Lee, K. et al. Magnetic Order and Symmetry in the 2D Semiconductor CrSBr. *Nano Lett.* **21**, 3511-3517 (2021).

12. Zhang, X. et al. Room-temperature intrinsic ferromagnetism in epitaxial $CrTe_2$ ultrathin films. *Nat. Commun.* **12**, 2492 (2021).

13. Zhu, W. et al. Large and tunable magnetoresistance in van der Waals ferromagnet/semiconductor junctions. *Nat. Commun.* **14**, 5371 (2023).

14. Wang, Z. et al. Electric-field control of magnetism in a few-layered van der Waals ferromagnetic semiconductor. *Nat. Nanotechnol.* **13**, 554-559 (2018).

15. Liang, S. et al. Small-voltage multiferroic control of two-dimensional magnetic insulators. *Nat. Electron.* **6**, 199–205 (2023).

16. Wang, Z.-A. et al. Selectively Controlled Ferromagnets by Electric Fields in van der Waals Ferromagnetic Heterojunctions. *Nano Lett.* **23**, 710–717 (2023).

17. Cenker, J. et al. Reversible strain-induced magnetic phase transition in a van der Waals magnet. *Nat. Nanotechnol.* **17**, 256–261 (2022).

18. Li, T. et. al. Pressure-controlled interlayer magnetism in atomically thin $CrI_3$. *Nat. Mater.* **18**, 1303–1308 (2019).

19. Song, T. et. al. Switching 2D magnetic states via pressure tuning of layer stacking. *Nat. Mater.* **18**, 1298–1302 (2019).

20. Chen, Y., Samanta, K., Shahed, N.A. et al. Twist-assisted all-antiferromagnetic tunnel junction in the atomic limit. *Nature* **632**, 1045–1051 (2024).

21. Min, K. H., Lee, D. H., Choi, S. J. et al. Tunable spin injection and detection across a van der Waals interface. *Nat. Mater.* **21**, 1144–1149 (2022).

22. Yang, H. et. al. Two-dimensional materials prospects for non-volatile spintronic memories. *Nature* **606**, 663–673 (2022).

23. Dong, S., Xiang, H., & Dagotto, E. Magnetoelectricity in multiferroics: a theoretical perspective. *Nat. Sci. Rev.* **6**, 629–641 (2019).

24. Jiang, S., Shan, J., & Mak, K. F. Electric-field switching of two-dimensional van der Waals magnets. *Nat. Mater.* **17**, 406–410 (2018).

25. Huang, B. et al. Electrical control of 2D magnetism in bilayer $CrI_3$. *Nat. Nanotechnol.* **13**, 544-548 (2018).

26. Song, T. et al. Voltage Control of a van der Waals Spin-Filter Magnetic Tunnel Junction. *Nano Lett.* **19**, 915–920(2019).

27. Lu, Y., Wang, H., Wang, L., & Yang, L. Mechanism of carrier doping induced magnetic phase transitions in two-dimensional materials. *Phys. Rev. B* **106**, 205403 (2022).

28. Ghosh, S., Stojić, N., & Binggeli, N. Overcoming the asymmetry of the electron and hole doping for magnetic transitions in bilayer $CrI_3$. *Nanoscale* **13**, 9391-9401(2021).

29. Huang, C. et al. Toward Intrinsic Room-Temperature Ferromagnetism in Two-Dimensional Semiconductors. *J. Am. Chem. Soc.* **140**, 11519 (2018).

30. Fei, Z. et al. Ferroelectric switching of a two-dimensional metal. *Nature* **560**, 336-339 (2018).

31. Wang, C. et al. Bethe-Slater-curve-like behavior and interlayer spin-exchange coupling mechanisms in two-dimensional magnetic bilayers. *Phys. Rev. B* **102**, 020402 (2020).

32. Sivadas, N., Okamoto, S., Xu, X., Fennie, C. J., & Xiao, D. Stacking-Dependent Magnetism in Bilayer $CrI_3$. *Nano Lett.* **18**, 7658–7664 (2018).

33. Yang, S. et al. Controlling the 2D Magnetism of $CrBr_3$ by van der Waals Stacking Engineering. *J. Am. Chem. Soc.* **145**, 28184–28190 (2023).

34. Liu, N. et al. Intralayer strain tuned interlayer magnetism in bilayer CrSBr. *Phys. Rev. B* **109**, 214422 (2024).

35. Kresse, G., & Hafner, J. Ab initio molecular dynamics for liquid metals. *Phys. Rev. B* **47**, 558 (1993).
36. Perdew, J. P., Burke, K., & Ernzerhof, M. Generalized Gradient Approximation Made Simple. *Phys. Rev. Lett.* **77**, 3865 (1996).
37. Dudarev, S. L., Botton, G. A., Savrasov, S. Y., Humphreys, C. J., & Sutton, A. P. Electron-energy-loss spectra and the structural stability of nickel oxide: An LSDA+U study. *Phys. Rev. B* **57**, 1505 (1998).
38. Blöchl, P. E. Projector augmented-wave method. *Phys. Rev. B* **50**, 17953 (1994).
39. Monkhorst, H. J., & Pack, J. D. Special points for Brillouin-zone integrations. *Phys. Rev. B* **13**, 5188 (1976).
40. Grimme, S. Semiempirical GGA-type density functional constructed with a long-range dispersion correction. *Comput. Chem.* **27**, 1787 (2006).
41. Makov, G., & Payne, M. C. Periodic boundary conditions in ab initio calculations. *Phys. Rev. B* **51**, 4014 (1995).

**Acknowledgements**

This work is supported by the Ministry of Science and Technology of the People´s Republic of China (No. 2022YFA1402901, 2022YFA1405100), by the NSFC (T2125004, 12274227, 12241405, 12004183), by the Fundamental Research Funds for the Central Universities (No. 30921011214, No. 30920041115), and by the Funding of NJUST (No. TSXK2022D002). C.H. and E.K. acknowledge the support from the Tianjing Supercomputer Centre and Shanghai Supercomputer Center.

**Author contributions**

E.K., C.H. and K.W. conceived the idea and supervised the project. C.H. performed calculations and data analysis. E.K., C.H. and K.W. co-wrote the paper. All authors discussed the results and commented on the manuscript at all stages.

## Additional information

**Supplementary Information** accompanies this paper at https://xxx, which includes details of tight-binding dimer model analysis, stacking patterns, electronic structures and MLWFs results for CrI$_3$/MnSe$_2$, magnetoelectric properties for other vdW magnetic bilayers and electric field effects on the magnetism of vdW magnetic heterobilayers.

**Competing financial interests:** The authors declare no competing financial interests.